\documentclass{hcj}
\usepackage{mdwmath}
\usepackage{bbding}
\usepackage{pifont}
\usepackage{wasysym}
\usepackage{amssymb}
\usepackage{mdwtab}
\usepackage{multirow}

\journalyear{2016}
\journalvolume{x}
\journalissue{x}
\journalpages{x-y}
\articledoi{xyz}
\journalcopyright{\copyright{} 2016,  xyz \& xyz. }

\title{Design Guidelines for the User-Centred Collaborative Citizen Science Platforms}
\author{Poonam Yadav\affil{Imperial College London}
       \and John Darlington\affil{Imperial College London}}
\authorrunning{P. Yadav, and J. Darlington}

\begin{document}

\maketitle

\begin{abstract}
Online Citizen Science platforms are good examples of socio-technical systems where technology-enabled interactions occur between scientists and the general public (volunteers). Citizen Science platforms usually host multiple Citizen Science projects, and allow volunteers to choose the ones to participate in. Recent work in the area has demonstrated a positive feedback loop between participation and learning and creativity in Citizen Science projects, which is one of the motivating factors both for scientists and the volunteers. This emphasises the importance of creating successful Citizen Science platforms, which support this feedback process, and enable  enhanced learning and creativity to occur through knowledge sharing and diverse participation. In this paper, we discuss how scientists' and volunteers' motivation and participation influence the design of Citizen Science platforms.  We present our summary as guidelines for designing these platforms as user-inspired socio-technical systems. We also present the case-studies on popular Citizen Science platforms, including our own CitizenGrid platform, developed as part of the CCL EU project, as well as Zooniverse, World Community Grid, CrowdCrafting and EpiCollect+ to see how closely these platforms follow our proposed guidelines and how these may be further improved to incorporate the creativity enabled by the collective knowledge sharing.
\end{abstract}

\section{Introduction}
In recent years, Citizen Science (CS)  has opened new territories of scientific collaboration by involving the general public. Internet-based Citizen Science projects attract volunteer  participants from diverse communities and benefit both scientists and volunteers. Due to the potential benefits, new projects are being regularly created. This invokes a need for the online Citizen Science platforms that can host different projects and allow volunteers to choose and participate. Examples of such platforms include CitizenGrid~\citep{Yadav2015CitizenGrid,Yadav2016CitizenGrid} that is developed as part of the CCL EU project, Zooniverse~\citep{Zooniverse} and others.  It is interesting to investigate what are the other benefits these platforms bring to both scientists and volunteers from a motivational perspective. Therefore, in this paper, we analyse both scientist's  and volunteer's participation cycles in a Citizen Science project, starting from their initial motivation. We present both scientists' and volunteers' participation steps in Section~\ref{Users}. In Section~\ref{Guidelines} we present the design guidelines for a user-centred Socio-technical collaborative Citizen Science platform, based on our understanding gained from the analysis. In Section~\ref{CaseStudies}, we study the already existing Citizen Science platforms with respect to the proposed guidelines to see how closely the platforms follow the guidelines and discuss how these may be further improved to incorporate the creativity based on collective knowledge sharing.  In Section~\ref{RelatedWork}, we present other similar online platforms that are used for  listing the Citizen Science projects and present the conclusions and a summary in Section~\ref{Conclusions}.  

\section{Users' Participation}\label{Users}
Citizen Science platforms are collaborative online spaces where scientific projects can be hosted to allow scientists to inform about their research and invite participants to contribute to it, while the general public (citizen scientists/ volunteers) can select and participate in the projects that fit their interests.  

Recent work in the area has revealed a positive feedback loop between participation and learning and creativity in Citizen Science projects, which is one of the motivating factors both for scientists and the volunteers to participate in the Citizen Science projects interactively. This emphasises the importance of creating successful Citizen Science platforms, which support this feedback process, and enable enhanced learning and creativity to take place both for the scientists and  the volunteers.   

In the context of this paper, the scientists seek through CS projects, various forms of public contribution to solve scientific problems they are working on. The volunteers are the general public who voluntarily contribute to these CS projects. In order to derive platform design guidelines that would take both types of users into consideration, we follow the user-centred design approach by analysing both the participation of the scientists and the volunteers in any type of Citizen Science project.

\subsection{Analysis of  Scientists' Participation Cycle}
\begin{figure}[h]
 \centering
 \includegraphics[width=90mm,height=50mm]{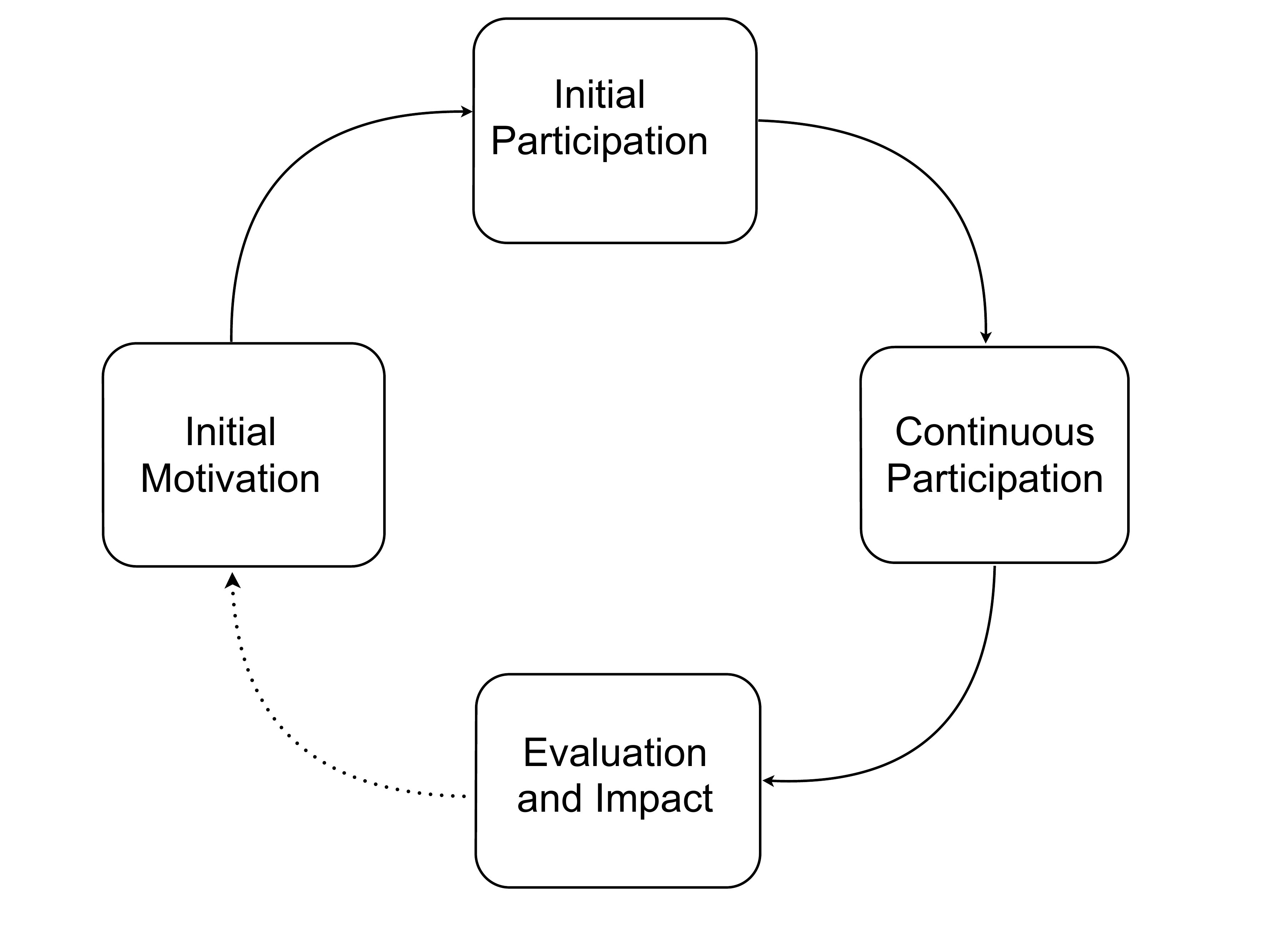} 
\caption{Scientist's Participation Cycle in a Citizen Science Project.}
\label{fig:APC} 
\end{figure}

The Scientists' participation cycle as shown in Figure~\ref{fig:APC} consists of the following steps: motivation, initial participation, continuous participation and evaluation and impact. The initial motivation defines the Scientist's interest in creating a Citizen Science project or converting a science project to Citizen Science Project. Once they are motivated, they look for the available support tools and techniques for creating and hosting the project online~\citep{Yadav2016CSframeworks}. Scientists also need monitoring tools of the volunteers' participation in order to enhance project outreach and engagement. The other tools they require for continuous monitoring for volunteers participation to enhance project outreach and engagement. The Scientists' participation cycle as shown in Figure~\ref{fig:APC} consists of the following steps: motivation, initial participation, continuous participation and evaluation and impact; we describe each step here in detail.  The initial motivation defines the Scientist's interest in creating  or converting a science project to a Citizen Science Project. Once they are motivated, they look for the available support tools and techniques for creating and project online hosting. The other tools they require for continuous monitoring for volunteers participation to enhance project outreach and engagement. 

\subsubsection{Initial Motivations}
Scientists may have one or more motivations to be involved in the citizen science projects. Some of these motivations are: (1) seeking help (contribution) in their research work; (2) seeking involvement of general public on global research issues for public awareness; (3) promoting and supporting the scientific literacy; (4) promoting their own research work~\citep{Charlene2013, Zandstra2015, Nov2011, Tweddle2012}. There could be various others reasons that make scientists seek help from the general public.  The one reason could be the nature of the research problem, for example, if it requires diversity contribution from a large number of participants. The other reason could be the unaffordable cost of the resources that are required to conduct research (both human and computing resources).

\subsubsection{Initial Participation}
Once the scientist is motivated to start a new CCS project, there are some factors that decide whether they would like to go further or not. These factors are categorised as technical, financial and legal. The main technical factors are: (1) The availability of frameworks and design tools to help forge an online Citizen Science project out from the initial science project; (2) The availability of platforms for hosting and deployment of the CS projects; (3) The process of handling the security and trust issues, privacy, and data quality issues~\cite{WClab2014, FederalCrowdsourcing, Gellman}. The financial factors relate to the affordable costs (in terms of time, man-power and resources), whereas the legal factors consider the process involved in handling the intellectual properties (IP) rights, and privacy and security related legal issues.

\subsubsection{Continuous Participation}
Once the CS project is deployed online, there are a number of factors they consider to keep continuing their participation in the project.  These factors are: (1) The availability of tools for analysing the volunteer's participation in the project; (2) Available help and support for changing or updating the various components of  the project; (3) The flexibility in changing the project without disturbing the volunteers who are already participating in the project; (4) Possibilities of interactive discussions with the volunteers.

\subsubsection{Evaluation and Impact}
The metrics, which measures the impact of the project, are defined corresponding to the project goals and outcomes.  In general, a project requires some frameworks and tools for measuring the impact of the project qualitatively or quantitatively.  
The frameworks and tools are required to: (1) measure the user engagement and participation; (2) measure the learning and creativity in short and long term; (3) measure the volunteer's contribution; (4) measure the project's outreach and publicity; (5) measure the project's comparative performance.

\subsection{Analysis of  Volunteers' Participation Cycle}
\begin{figure}[h]
 \centering
 \includegraphics[width=70mm,height=50mm]{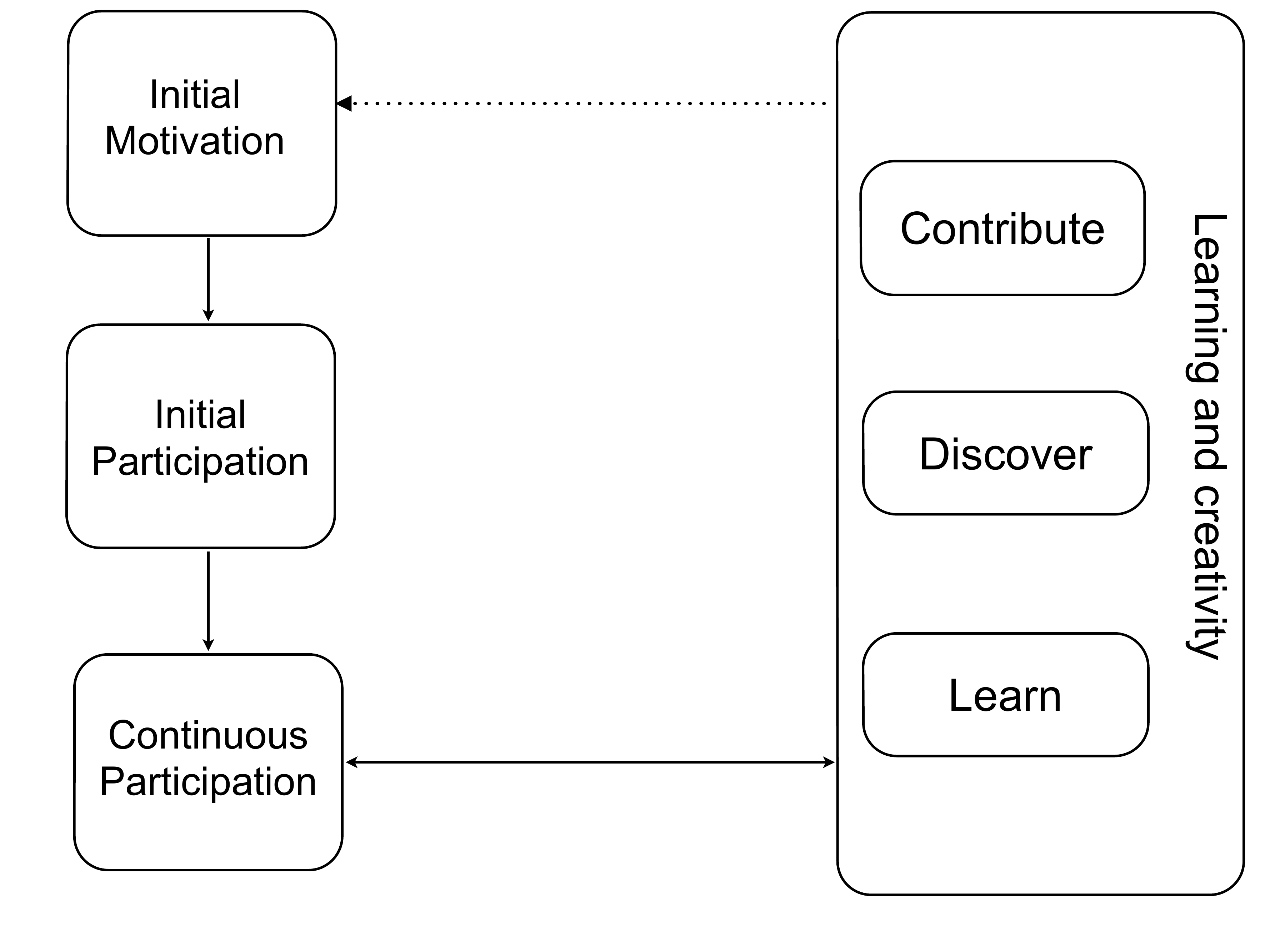} 
\caption{Volunteer's Participation Cycle in a Citizen Cyber Science Project.}
\label{fig:AUC} 
\end{figure}

In recent years, a number of research studies have been conducted to understand the volunteers involvement in Citizen Science~\citep{Charlene2013, Zandstra2015, Nov2011, Tweddle2012}.  Figure~\ref{fig:AUC}  shows the volunteers participation cycle in a Citizen Science project, starting from initial motivation to continuous participation.

\subsubsection{Initial Motivation}
The well-known motivations~\citep{Tweddle2012, Nov2011} are: (1) willingness and desire to contribute in science; (2) desire to learn science by involving in Citizen Science; (3) for fun and enjoyment through game playing or other interactive project participation. 

\subsubsection{Initial Participation}
Once volunteers are interested in contributing in a Citizen Science project, their initial participation are based on the following factors: (1) easy discovery of interesting Citizen Science projects; (2) ease and simplicity of launching the project; (3) quick, easy and secure project's initial setup.

\subsubsection{Continuous Participation}
The volunteers continuous participation depends on their expectations that come from their initial motivations. The three main points that influence volunteer's long-term engagement meant are: (1) volunteers receive feedback and appreciation for their participation~\citep{Charlene2014}; (2) volunteers can measure or observe what they are learning while participating in the project; (3) if the project meets volunteers' fun and enjoyment expectations~\citep{Jackson2015}.

\section{The Collaborative Platform's Design Guidelines}\label{Guidelines}
The goal of Citizen Science platforms is to bring together a number of Citizen Science projects for volunteers participation. The platforms could provide these different functionalities, such as (1) project portal where volunteers can find their favourite project by searching or selecting from dashboard, menus, etc.; (2) host complete project or project clients; (3) launch and run project servers. All Internet-based Citizen Science (Virtual Citizen science) projects are not similar and may fall into different project categories and have different deployment scenarios, for example volunteer computing and human-computation CS projects~\citep{Yadav2014Deployment, Yadav2016CSdeployment}. Moreover, Citizen Science platforms are also designed for either one or more project categories. Therefore, the design guidelines we present here are generic and are based on our experiences gained from building the CitizenGrid Platform and could be applied to any Citizen Science platform regardless the project categories the platform supports.  
\begin{itemize}
\item [1.] \textbf{Cost Effective}: This is an important consideration for a setting up Citizen Science projects, since most are started by an academic or a public institution with a limited initial grants. The use of open source technologies in the platform/project development or free to use services is an important consideration for the platform usability, having a wider scientific community in mind.
\item [2.] \textbf{Easy to Create Project}: The easy, simple and quick project creation and deployment process allows scientists to deploy and maintain their project themselves (DIY: do-it-yourself feature). 
\item [3.] \textbf{Multi-categories Projects}: If a Citizen Science platform supports multi-categories/types of projects, it would make it easier for the scientists to advertise, host, and maintain different types of Citizen Science projects at the same place. For example, this would allow flexibility to host and deploy volunteer thinking, volunteers computing and game-based Citizen Science projects. This feature helps volunteers to discover and participate in different types of projects on one platform.  
\item [4.] \textbf{Comparative Project Performance}: This feature provides scientists the comparative summary report of their project, including volunteers engagement and participation report, volunteers learning and creativity report, project overall impact and volunteer contribution report. 
\item [5.] \textbf{Easy Maintenance}: For long-term projects, it is a very important design consideration. The platform should allow easy project file updates - adding new files, deleting, synching new information without causing problems to participating volunteers.
\item [6.] \textbf{Security and Trust}: The platform should adhere security requirements that are required for the standard web platforms.
\item [7.] \textbf{Scientists and Volunteers Interaction}: This is important for users engagement and long-term participation. The interaction can be achieved by user forums, online videos for help and tutorials, real-time message exchange systems, etc.
\item [8.]\textbf{Simple Project Participation}: The volunteers who contribute in the Citizen Science projects advertised on the platforms have varied computing skills levels ranging from novice to expert. The devices they use to access the platforms are also heterogeneous in terms of hardware capabilities and software. Therefore, it is very necessary that platform make the project simple and easily accessible to them.
\end{itemize}

\section{Case Studies}\label{CaseStudies}
In this section we analyze already existing platforms World Community Grid~\citep{WCG2015}, CitizenGrid~\citep{Yadav2015CitizenGrid,CitizenGrid,Citizengrid2014},  Zooniverse~\citep{Zooniverse, Tinati2015}, Epicollect+~\citep{Epicollect,Aanensen2014}, and CrowdCrafting~\citep{Crowdcrafting}. All platforms are free to use and allow easy project creation.   CitizenGrid, Epicollect and  Zooniverse, CrowdCrafting are open source  and  supports do-it-yourself (DIY) new project creation process.  In this section, we analyse all these platforms with respect to the design guidelines  we presented above and present summary in Table~\ref{CSplatforms}.

\textbf{CitizenGrid:}  CitizenGrid allows do-it-yourself hosting of the project and simplify the process of hosting and deployment of the project by supporting Virtual Machine based client distribution. 
The CitizenGrid allows flexibility to use any of the task management and distribution frameworks from the list:  Copilot~\citep{Copilot}, LiveQ~\citep{LiveQ}, or BOINC~\citep{Boinc} and 
allows project discovery  and search mechanism by free text search  for volunteers. On CitizenGrid interface volunteers can see their participation history and scientists can see  projects' comparative performance, and  easily maintain their project by  changing the project new files etc.. However, in current version  any change in the client part requires volunteers to restart the client again.  CitizenGrid team consists of four members and have hosted 4 real projects, e.g., RedWire~\cite{Redwire2015} and Cern Virtual Atom Smasher Game~\cite{VAS2015a, Yadav2016}  and advertised nearly 50 already existing  volunteer computing and thinking projects. CitizenGrid doesn't maintain social network account but hold a YouTube channel and  a regular meet up event  for one-to-one interaction and have nearly 25 registered users. 

\textbf{World Community Grid:} World Community Grid platform hosts volunteer computing (VC) projects e.g., Uncovering Genome Mysteries~\citep{UGM} or Mapping Cancer Markers~\citep{MCM2013} and CERN Test4Theory~\citep{Test4Theory}. The World Community Grid uses BOINC (Boinc, 2015) middleware framework for the projects task management and distribution. 
World Community Grid's project support-team provides full support for hosting and maintaining the scientist volunteer computing project, if the platform team  approves the project. The platform hosts 5 active projects  currently with 26	 projects in total.  World Community Grid  is very popular platform with social networking platforms followers (Facebook: 22196 and Twitter: 7.6 K  (on 28th December 2015) ).

\begin{figure}[h]
 \centering
 \includegraphics[width=100mm,height=60mm]{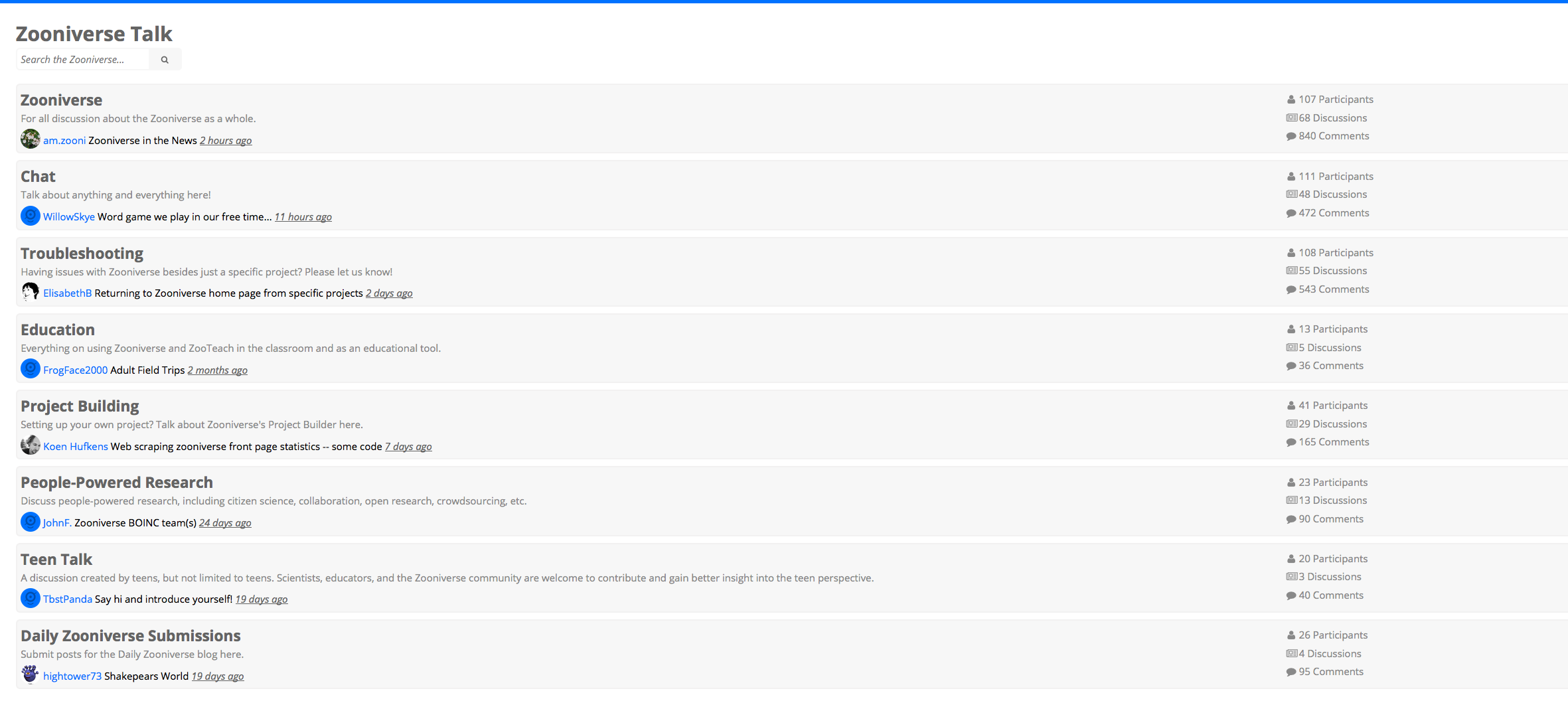} 
\caption{Zooniverse Interactive Forum (screenshot taken on 27th December 2015) ~\citep{Zooniverse} .}
\label{fig:zooniverse} 
\end{figure}

\textbf{Zooniverse:} Zooniverse platform allows do-it-yourself hosting of volunteer cognition and thinking (VT) projects. In volunteer cognition and thinking projects, where volunteers  only need to use their time, cognitive power and knowledge to analyse data. Zooniverse is currently hosting 42 projects (on 28th December 2015) with a very user friendly interface, however,  there is no option for searching a project by name or category.  The projects comparative performances are not available to the project scientists, however, information is made available through  research publications~\citep{Tinati2015}. The Zooniverse development and maintenance team consists of nearly 20 members and funded by a number of research grants. Zooniverse is very popular platform with social networking platforms followers (Facebook: 24551 and Twitter: 13.5 K  (on 28th December 2015) )  and maintains active blogs $-$ \textit{daily Zooniverse} for a quick daily update and Zooniverse blogs.  Addition to this, the platform maintenance team maintains an interactive talk forum as shown in Figure~\ref{fig:zooniverse} for volunteers' interactions. This is very important feature which motivates volunteers to stay connected with the project. The detail analysis of the interaction has been presented in the research articles (Tinati, 2014). The volunteer participation in volunteer cognition projects is marked as creative, learning oriented and interactive~\citep{Charlene2014}. \\

\textbf{Epicollect+:}
The Epicollect+~\citep{Epicollect,Aanensen2014} platform allows data collection and aggregation platform. It is supported by the mobile client app, which makes it easy for volunteers to support and participate in the data collection projects. It allows scientists to setup the do-it-yourself data collection projects, for example, photos and text based survey, etc.  Epicollect+ have two components: a mobile app and a web server app.  The EpiCollect+ mobile app allows scientists to load a single or multiple projects and provides the interface for volunteers to gather the data.  All data can be subsequently synched with a central server and, furthermore, data can be retrieved onto the mobile devices from the central server and viewed as tables or maps. The web server app  provides facility of hosting the server for project data, by providing server software and instructions for setting up a database and web application for scientists to house and view the data collected by any number of mobile devices.

\textbf{CrowdCrafting:}
CrowdCrafting hosts human computation projects as well as data collection projects such as image classification, transcription, geocoding and  uses Epicollect+ for the mobile clients. CrowdCrafting has a very appealing interface with easy project discovery in seven categories: featured, social, art, humanities, biology, economics, and science and allows do-it-yourself (DIY) project creation process.

\begin{table}[h]
\centering
\caption{Citizen Science Online Platforms}
\label{CSplatforms}
\scalebox{0.55}{
\begin{tabular}{| c | c | c | c | c | c | c | c | c | c | c |}
\hline 
 \textbf{PLATFORMS} &   \multicolumn{10} {|c|} {\textbf{GUIDELINES}} \\\hline
  &  \multicolumn{2} {|c|} {Cost Effective} & \multicolumn{2} {|c|} {Easy Project Creation} & Multi-Categories  & Comparative  & Easy  & Security  \& & Interaction & Simple   \\\cline{3-6}
 & Free-to-use& Open-source& General& DIY&Projects &Performance  &Maintenance &Trust  &&  Participation   \\\hline
CitizenGrid &  $\checkmark$& $\checkmark$&$\checkmark$& $\checkmark$& VC\&VT& -- & $\checkmark$& $\checkmark$& -- &$\checkmark$ \\\hline
World Community &   $\checkmark$& $\times$&$\checkmark$& $\times$& VC& -- & $\checkmark$& $\checkmark$& -- &$\checkmark$ \\\hline
Zooniverse  &  $\checkmark$ & $\checkmark$&$\checkmark$& $\checkmark$& VT& -- & $\checkmark$& $\checkmark$& $\checkmark$  &$\checkmark$ \\\hline
Epicollect  & $\checkmark$ & $\checkmark$& $\checkmark$& $\checkmark$& DC& -- & $\checkmark$ & $\checkmark$& -- &$\checkmark$ \\\hline 
CrowdCrafting & $\checkmark$ & $\checkmark$& $\checkmark$& $\checkmark$& DC& -- & $\checkmark$ & $\checkmark$& -- &$\checkmark$ \\\hline 

\end{tabular} }
\scalebox{0.7}{
-- : Partially Supported  
}
\end{table}

\section{Related Work} \label{RelatedWork}
Design guidelines for online platform play central  role in human-computer-interface (HCI) research. Apart from the platform discussed in previous section, in recent year,  there are a number of Citizen Science Portals such as  Scistarter~\citep{SciStarter2014}, Citizen Science Search~\citep{CSSearch2014}, and  Wikipedia~\citep{Wikipedia} are developed. These portals  list a number of Citizen Science projects for easy project discovery. SciStarter is a web-portal for Citizen Science projects that maintains a catalogue of nearly 700 projects and provides different search criteria such as outdoor/indoor projects, featured projects and projects suitable for students and children, etc. SciStarter itself does not host projects, instead, direct volunteers to the individual project websites where potential volunteers can find out more about a project and participate. Citizen Science Search website offers "keyword" based text search and list only a limited number of selected projects. Wikipedia lists and displays many Citizen Science application and projects with their brief description along with project links.
 
\section{Conclusions}\label{Conclusions}
In this paper,  we presented the design guidelines for user-inspired Citizen Science collaborative platforms by analysing the participation steps of scientists and volunteers in a Citizen Science project. We discussed how scientists' and volunteers' motivation and participation influence the design of these platforms.  We also presented the case-studies on popular Citizen Science platforms such as Zooniverse, CitizenGrid, World Community Grid, EpiCollect  and CrowdCrafting to see how closely these platforms matches our proposed guidelines. These case studies are helpful in understanding how closely these platforms follow our proposed guidelines and how these may be further improved to incorporate the creativity based on collective knowledge sharing.
\bibliography{hcj-citizengrid}

\end{document}